\documentclass[preprint]{elsarticle}

\usepackage{algorithm}      %For algorithm
\usepackage{algpseudocode}  %For algorithm
\usepackage{booktabs}       %For table
\usepackage{multirow}       %For table

\usepackage{amssymb}
\usepackage{amsmath}
\usepackage[pagewise]{lineno}
\usepackage{geometry}
\makeatletter
\def\ps@pprintTitle{%
	\let\@oddhead\@empty
	\let\@evenhead\@empty
	\def\@oddfoot{\reset@font\hfil}%
	\let\@evenfoot\@oddfoot
}	
\makeatother

\journal{Journal of Computational Physics}

\begin{document}
\begin{frontmatter}

\title{A CNN-based particle tracking method for large-scale fluid simulations with Lagrangian-Eulerian approaches}

\author{
  Xuan Luo\textsuperscript{a},
  Zichao Jiang\textsuperscript{a},
  Yi Zhang\textsuperscript{b},
  Qinghe Yao\textsuperscript{a}*\corref{cor1},
  Zhuolin Wang\textsuperscript{a}, 
  Gengchao Yang\textsuperscript{a}, 
  Bohua Huang\textsuperscript{a}
}

\affiliation{organization1={School of Aeronautics and Astronautics, Sun Yat-sen University},
            city={Guanzhou},
            postcode={510220}, 
            state={Guangdong},
            country={China}
            }
            
\affiliation{organization2={School of Mathematics and Computing Science, Guilin University of Electronic Technology},
            city={Guilin}, 
            postcode={541004}, 
            state={Guangxi}, 
            country={China}}

\cortext[cor1]{Corresponding author: Qinghe Yao (yaoqh@mail.sysu.edu.cn)}

\begin{abstract}
  
  A novel particle tracking method based on a convolutional neural network (CNN) is proposed to improve the efficiency of Lagrangian-Eulerian (L-E) approaches. Relying on the successive neighbor search (SNS) method for particle tracking, the L-E approaches face increasing computational and parallel overhead as simulations grow in scale. This issue arises primarily because the SNS method requires lengthy tracking paths, which incur intensive inter-processor communications. The proposed method, termed the CNN-SNS method, addresses this issue by approximating the spatial mapping between reference frames through the CNN. Initiating the SNS method from CNN predictions shortens the tracking paths without compromising accuracy and consequently achieves superior parallel scalability. Numerical tests demonstrate that the CNN-SNS method exhibits increasing computational advantages over the SNS method in large-scale, high-velocity flow fields. As the resolution and parallelization scale up, the CNN-SNS method achieves reductions of $95.8\%$ in tracking path length and $97.0\%$ in computational time.

\end{abstract}


\end{frontmatter}



\section{Introduction}
\label{sec:sec1}

The Lagrangian-Eulerian (L-E) approach denotes a suite of modeling methods wherein droplets or particles are represented in the Lagrangian reference frame, while the carrier-phase flow field is described in the Eulerian reference frame \citep{subramaniam2013lagrangian}. It offers notable advantages over traditional Euler methods in handling complex boundary problems. It has proven to be instrumental in simulations of various complex engineering problems in computational fluid dynamics, including multiphase flows \citep{subramaniam2013lagrangian, diggs2016evaluation}, multiscale flows \citep{wen2020atomization, li2023investigation, mira2023hpc}, and fluid-structure interactions \citep{kolahdouz2023sharp, mittal2008versatile, han2020eulerian, Zheng2021interface}.

However, the L-E approaches often encounter a L-E coupling problem during data exchanges between the two reference frames, necessitating the establishment of a spatial mapping between them \citep{goron2023simulation, casadei2011strong, cheron2023coupled}. For instance, the Lagrangian-Galerkin (L-G) method \citep{karakashian1982galerkin}, a specific L-E approach recognized for its stability, combines a Galerkin finite element procedure with a special discretization of the material derivative along the trajectories of Lagrangian particles \citep{bermejo2023new, massarotti1998characteristic}. Coupling the Lagrangian trajectories with the Eulerian model requires constructing the spatial mapping, which is achieved by identifying the host element of each particle within the Eulerian model \citep{bermejo2010semi, pironneau1982transport}. Similar coupling problems are also encountered in other related methods. For instance, the semi-Lagrangian method \citep{sawyer1963semi, xiu2001semi} and the characteristic-Galerkin method \citep{zienkiewicz1995split} couple fluid particles with Eulerian flow fields. Similarly, in the front-tracking method \citep{unverdi1992front} and the fluid-structure interaction algorithm \citep{casadei2011strong, zhang2022numerical, zhang2021optimized}, Eulerian interfacial flows are coupled with deforming phase Lagrangian boundaries \citep{muradoglu2006auxiliary}. Although this coupling problem is trivial in structured meshes where the host elements are sequentially arranged and can be quickly identified, it is more challenging in unstructured meshes where the spatial mappings become complicated, necessitating solution via particle tracking methods \citep{lohner1990vectorized, wang2022gpu, gassmoller2018flexible}. As the tracking process becomes more computationally expensive with an increasing resolution of the Eulerian model, an effective particle tracking method is crucial for the efficiency of the L-E approaches.

Various particle tracking methods have been proposed in earlier studies. The auxiliary structured grid (ASG) method \citep{muradoglu2006auxiliary} overlays the original unstructured Eulerian model with a Cartesian grid, wherein particles can be readily assigned to specific Cartesian elements, thereby narrowing the tracking domain. However, the effectiveness of the Cartesian grid is limited in meshes with large-scale element size variations. The hierarchy tree (HT) method was developed to address this issue, employing a tree-like hierarchy of Cartesian meshes to accommodate varying element sizes \citep{giraldo2003strong}. Still, both methods incur considerable storage overhead when applied in high-resolution Eulerian models. The successive neighbor search (SNS) method is more commonly used due to its stability and efficiency \citep{allievi1997generalized}. Its tracking initiates from the host element of particles before movement, determining the tracking direction via barycentric coordinates within each element along the path until reaching targets. While in the regular Eulerian model, the target can be achieved iteratively through finite steps, more iterations are required as the resolution of the Eulerian model grows, leading to inefficiency in the tracking process for large-scale simulations. 

In addition, the computational cost of the SNS method incurred by the lengthy tracking path is further exacerbated in parallel frameworks. A standard SNS parallelization strategy is to divide the Eulerian model into multiple subdomains and assign each subdomain to separate processes \citep{kaludercic2004parallelisation, thari2021parallel, dubey2011parallel}. In this Eulerian-based parallelization, Lagrangian data need to be communicated whenever the tracking path crosses the boundaries of the subdomains \citep{beaudoin2007efficient, gimenez2014evaluating}. Consequently, as the level of parallelization increases and the boundaries become more compact, the tracking path becomes increasingly fragmented, resulting in a more significant communication burden and poor parallel scalability. Primarily, the computational inefficiency of the SNS method stems from the inaccuracy of the tracking origin, which is a simple and reasonable initial guess but still far away from the target in high-resolution Eulerian models \citep{bermejo2010semi}. This reason, in turn, suggests that the efficiency of the SNS method can be improved by replacing the initial guess with a more precise prediction.

Essentially, predicting the tracking target relies on approximating the spatial mapping, rendering neural networks particularly effective due to their nonlinear mapping capability. In recent years, Neural networks have been widely applied to computational fluid dynamics \citep{brunton2020machine, cai2021physics, jiang2023neural, jiang2024neural}, among which the convolutional neural networks (CNNs) are particularly noted for their ability to capture spatial information \citep{portal2022cnn, chen2021compressed, renyu2024super}. These features make the CNNs ideally suited to assist the SNS method. Furthermore, numerous similar particles are repeatedly mapped into the Eulerian model across all time steps in large-scale L-E simulations, particularly in steady and periodic flows. These repetitive, costly operations can not only offer ample training data but also be approximated by a well-trained CNN. Nevertheless, few studies in the existing literature have explored the use of CNNs or other neural network-based methods for particle tracking.

Therefore, we propose an optimized particle tracking method, termed the CNN-SNS method, integrating the SNS with a CNN. In this method, data from both reference frames of the L-E solver are first preprocessed by the ASG method for simplification. Subsequently, the preprocessed data is fed into the CNN, where convolutional and pooling layers capture the spatial mapping and output the tracking prediction. Initiating the SNS method from the prediction shortens the tracking path considerably, resulting in reduced computational consumption. Furthermore, the intersections between the shortened tracking paths and subdomain boundaries are consequently reduced, thereby achieving superior parallel in the CNN-SNS method. Overall, the CNN does not serve as a surrogate model. Instead, it approximates the tracking process to optimize the initial guess for the SNS method. Such integration preserves accuracy and circumvents the typical lack of interpretability associated with neural networks.

In this paper, we evaluate the capabilities of the CNN-SNS method by applying it to an L-G solver. The remainder of this paper is organized as follows. Section 2 details the particle tracking requirement of the L-G method and outlines the algorithms for both the SNS and the CNN-SNS methods. Section 3 presents the results of benchmark simulations, including a lid-driven cavity flow and a flow around a sphere. Finally, conclusions are drawn in Section 4.

\section{Methodology}
\label{sec2}

%% Use \subsection commands to start a subsection.
\subsection{Governing equation and discretization in the L-G method}
\label{sec:subsec1}

Without loss of generality, we briefly introduce the problem in the following setting. The dimensionless form three-dimensional incompressible Navier-Stokes equations in the time-space domain $\Omega\times\left(0, T\right]$ are written as
\begin{equation}
  \label{equ:1}
  \begin{aligned}
  	&\dfrac{\partial \boldsymbol{u}}{\partial t} + (\boldsymbol{u} \cdot \nabla) \boldsymbol{u} -\frac{ 1}{\mathrm{Re}} \Delta \boldsymbol{u}+\nabla p =  \boldsymbol{f},   \\
  	&\nabla \cdot \boldsymbol{u} = \boldsymbol{0}, 
  \end{aligned}
\end{equation}
where $\Omega$ is a connected bounded subset of $\mathbb{R}^3$ with a Lipschitz continuous boundary $\partial\Omega$, $\boldsymbol{u}$ is velocity, $\mathrm{Re}$ is the Reynolds number, $p$ is pressure, and $\boldsymbol{f}$ is external body force. The problem is closed with an initial condition and a boundary condition,
\begin{equation}
  \begin{aligned}
  &\boldsymbol{u}(\boldsymbol{x}, 0) = \boldsymbol{u}_0 && \quad \text{in }\Omega, \\
  &\boldsymbol{u}(\boldsymbol{x}, t) = \boldsymbol{0} && \quad \text{on }\partial \Omega\times\left(0, T\right].\nonumber
   \end{aligned}
\end{equation}
In a finite element setting, a weak form of \eqref{equ:1} is given as: Seek $\left(\boldsymbol{u},p\right)\in\left[H_{0}^1(\Omega)\right]^3 \times L^2(\Omega)$, such that 
\begin{equation}
  \label{equ:2}
  \begin{aligned}
  &\left( \dfrac{\mathrm{D} \boldsymbol{u}}{\mathrm{D} t}, \boldsymbol{v} \right)_{\Omega} + \dfrac{1}{\mathrm{Re}}\left(\nabla \boldsymbol{u}, \nabla \boldsymbol{v}\right)_{\Omega} - \left(\nabla \cdot \boldsymbol{v}, p\right)_{\Omega} = \left( \boldsymbol{f}, \boldsymbol{v} \right)_{\Omega} && \forall \boldsymbol{v} \in \left[H_{0}^1(\Omega)\right]^3, \\
  &\left(\nabla \cdot \boldsymbol{u}, q\right)_{\Omega} = 0 && \forall q \in L^2(\Omega),
\end{aligned}
\end{equation}
where $\dfrac{\mathrm{D} \boldsymbol{u}}{\mathrm{D} t} = \dfrac{\partial \boldsymbol{u}}{\partial t} + \left(\boldsymbol{u} \cdot \nabla\right) \boldsymbol{u}$.

Solving \eqref{equ:2} with the Galerkin finite element requires the integration of the nonlinear term $\left( \frac{\mathrm{D} \boldsymbol{u}}{\mathrm{D} t}, \boldsymbol{v} \right)_{\Omega}$ over elements. The essence of the L-G method lies in approximating the integration by discretizing $\frac{\mathrm{D} \boldsymbol{u}}{\mathrm{D} t}$ along the Lagrangian particle trajectories \citep{suli1988convergence}. Representing the particle located at position $\boldsymbol{x}$ at time $t=t_n$ as $\boldsymbol{X}(\boldsymbol{x}, t_n; t)$, then the first-order discretization of $\frac{\mathrm{D} \boldsymbol{u}}{\mathrm{D} t}$ can be written as
\begin{equation}
  \label{equ:3}
  \begin{aligned}
  \dfrac{\mathrm{D} \boldsymbol{u}}{\mathrm{D} t}\left(\boldsymbol{x}, t_n\right) =\frac{\boldsymbol{u}\left(\boldsymbol{X}\left(\boldsymbol{x}, t_n ; t_n\right), t_n\right)-\boldsymbol{u}\left(\boldsymbol{X}\left(\boldsymbol{x}, t_n ; t_{n-1}\right), t_{n-1}\right)}{\Delta t}+O(\Delta t).
  \end{aligned}
\end{equation}
Consider the example of a tetrahedral mesh with finite elements $\left\{K_l \right\}^{N_e}_{l=1}$, where $N_e$ denotes the number of elements in the Eulerian model. In any element $K_m \in \left\{K_l \right\}^{N_e}_{l=1}$, the integration of \eqref{equ:3} involves 
\begin{equation}
  \int\limits_{\substack{K_m}} \boldsymbol{u}\left(\boldsymbol{X}\left(\boldsymbol{x}, t_n ; t_{n-1}\right), t_{n-1}\right) \mathrm{~d} \boldsymbol{x},\nonumber
\end{equation}
which can be approximated by \citep{yao2010balancing},
\begin{equation}
  \label{equ:4}
  \int\limits_{\substack{K_m}} \sum_{i=1}^3 \sum_{j=1}^4\left(\boldsymbol{u}\left(\boldsymbol{X}\left(\boldsymbol{n}_j, t_n ; t_{n-1}\right), t_{n-1}\right) \varphi_j \boldsymbol{e}\right) \varphi_i \mathrm{~d} \boldsymbol{x},
\end{equation}
where $\left\{\boldsymbol{n}_j\right\}_{j=1}^4$ are the nodes of $K_m$, and $\boldsymbol{e}$ is the component of the canonical basis, and $\varphi_j$ is defined such that
\begin{equation}
  \varphi_i\left(\boldsymbol{n}_j\right)= \begin{cases}1 & (i=j), \\
   0 & (i \neq j).\end{cases}\nonumber
\end{equation}
To solve \eqref{equ:4}, $\boldsymbol{u}\left(\boldsymbol{X}\left(\boldsymbol{n}_j, t_n ; t_{n-1}\right), t_{n-1}\right)$ is approximated by interpolation within the host element containing $\boldsymbol{X}\left(\boldsymbol{n}_j, t_n ; t_{n-1}\right)$, which has yet to be determined. Let $\boldsymbol{X}\left(\boldsymbol{n}_j, t_n ; t_{n-1}\right)$ be denoted as $\boldsymbol{x}_p=[x_p,y_p,z_p]$, the remaining challenge lies in identifying the host element of $\boldsymbol{x}_p$, denoted as $K_t$, in the Eulerian model.

Identifying $K_t$ is straightforward in uniform and structured meshes only, in which the mapping between $\left\{K_l\right\}_{l=1}^{N_e}$ and $\boldsymbol{x}_p$ is linear. Otherwise, high-efficiency particle tracking methods are necessary, motivating the development of the CNN-SNS method.

\subsection{The SNS method}
\label{subsec2}

To identify $K_t$, the SNS method initiates the search from $K_m$ and determines the tracking direction by the barycentric coordinates of $\boldsymbol{x}_p$ within each $K_l$ along the tracking path.

Within each $K_l$, the particle's barycentric coordinates $\lambda_i^l(\boldsymbol{x}_p)(i=1,\cdots,4)$ are defined by
\begin{equation}
  \lambda_i^l(\boldsymbol{x}_p) = |S_i||S_0|^{-1} \quad (i = 1, \cdots, 4)\nonumber
\end{equation}
where the matrix $S_0$ is defined as
\begin{equation}
  S_0=\left[\begin{array}{ccc}
  1 & \cdots & 1 \\
  x_l^1 & \cdots & x_l^4 \\
  y_l^1 & \cdots & y_l^4 \\
  z_l^1 & \cdots & z_l^4
  \end{array}\right]\nonumber
\end{equation}
and the matrix $S_i$ is derived by replacing the $i$-th column of $S_0$ with $\left[x_p, y_p, z_p\right]^T$, and $\left\{x_l^i, y_l^i, z_l^i\right\}_{i=1}^4$ represent the coordinates of $\left\{\boldsymbol{n}_j\right\}_{j=1}^4$. The neighboring elements of $K_l$ are defined by
\begin{equation}
  \left\{m_1^l, \cdots , m_4^l\right\} \subset\left\{\left\{K_l\right\}_{l=1}^{N_e} \cup\{-1\}\right\},\nonumber
\end{equation}
where $\left\{m_1^l, \cdots ,m_4^l\right\}$ are the four neighboring elements of $K_l$, and $\{-1\}$ represents the absence of neighboring elements, indicating that $K_l$ lies on the the boundary of the Eulerian model. 

It can be proven that $\boldsymbol{x}_p$ located in $K_l$ if and only if $\lambda_i^l \geq 0$ for all $i \in \left\lbrace 1, \cdots, 4\right\rbrace$. Otherwise, the tracking direction from $K_l$ to $\boldsymbol{x}_p$ can be determined by finding $j \in \left\lbrace1, \cdots, 4\right\rbrace$ that satisfies
\begin{equation}
  \label{equ:5}
  \lambda_j^l(\boldsymbol{x}_p)=\operatorname{min}\left\{\lambda_i^l\left(\boldsymbol{x}_p\right)\right\}_{i=1}^4.
\end{equation}
Therefore, among $\left\{m_1^l, \cdots , m_4^l\right\}$, $m_j^l$ is the element that is closest to $\boldsymbol{x}_p$, and is thus selected as the next element. Through iterating the above process, the SNS method approaches $\boldsymbol{x}_p$ element-by-element. Alternatively, the tracking iteration may lead to the boundary of the Eulerian model, indicating that $\boldsymbol{x}_p$ lies outside the domain of $\left\{K_l\right\}_{l=1}^{N_e}$. Overall, the SNS method can be described as
\begin{equation}
  \label{equ:6}
  K_t=\operatorname{SNS}\left(\boldsymbol{x}_p, K_m,\left\{K_l\right\}_{l=1}^{N_e},\left\{m_1^l, \cdots m_4^l\right\}_{l=1}^{N_e}\right),
\end{equation}
with the process shown in Algorithm~\ref{alg:SNS}

\begin{algorithm}[H]
  \caption{The SNS method}\label{alg:SNS}
  \begin{algorithmic}[1]
  \State \textbf{Input:} $\boldsymbol{x}_p, K_m, \{K_l\}_{l=1}^{N_e}, \{m_1^l, \dots, m_4^l\}_{l=1}^{N_e}$
  \State \textbf{Output:} $K_t$

  \State $K_t \gets K_m$
  \State $I_t \gets 0$
  \While{true}
      \State $I_t \gets I_t + 1$
      \State Calculate $\boldsymbol{x}_p$'s barycentric coordinates $\lambda_i^l(\boldsymbol{x}_p)(i=1,\cdots,4)$ in $K_t$
      \If{$\lambda_i^l \geq 0$ for all $i \in 1, \cdots, 4$}
          \State \textbf{break}
      \EndIf
      \State Find $j \in \{1, \dots, 4\}$ that satisfy \eqref{equ:5}
      \If{$m_j^t \neq -1$}
          \State $K_t \gets m_j^l$
      \Else
          \State \textbf{break}
      \EndIf
  \EndWhile
  \State \Return $K_t$
  \end{algorithmic}
\end{algorithm}

In Algorithm~\ref{alg:SNS}, $I_t$ denotes the number of iterations the SNS method takes to reach $K_t$, which is positively correlated with both the density of elements and the distance between $K_t$ and $K_m$. In the L-G method, the density of elements is proportional to $N_e$, and the distance between $K_t$ and $K_m$ is determined by $|\boldsymbol{u}|\Delta t$. Therefore, the SNS method would incur a significant iteration overhead in L-G simulations with high resolution and velocity.

\subsection{The CNN-SNS tracking method}
\label{subsec3}

\begin{figure}[H]%% placement specifier
  \centering%% For centre alignment of image.
  \includegraphics[width=1\linewidth]{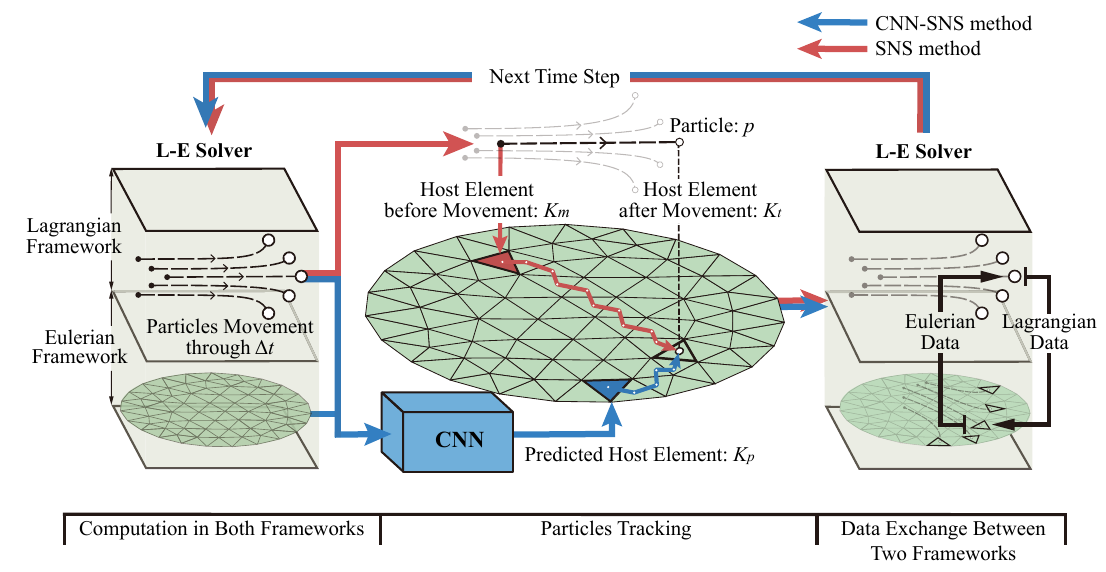}
  %% Use \caption command for figure caption and label.
  \caption{The solving process of the L-E solver in a time step, with the SNS and the CNN-SNS methods to solve the particle tracking problem.}\label{fig:Fig1}
\end{figure}

To reduce $I_t$ required by the SNS method, the CNN-SNS method shortens its tracking path by predicting $K_t$ through the CNN. The prediction, denoted as $K_p$, replaces $K_m$ as the tracking origin of the SNS process, as illustrated in Figure~\ref{fig:Fig1}.

\begin{figure}[H]
  \centering
  \includegraphics[width=1\linewidth]{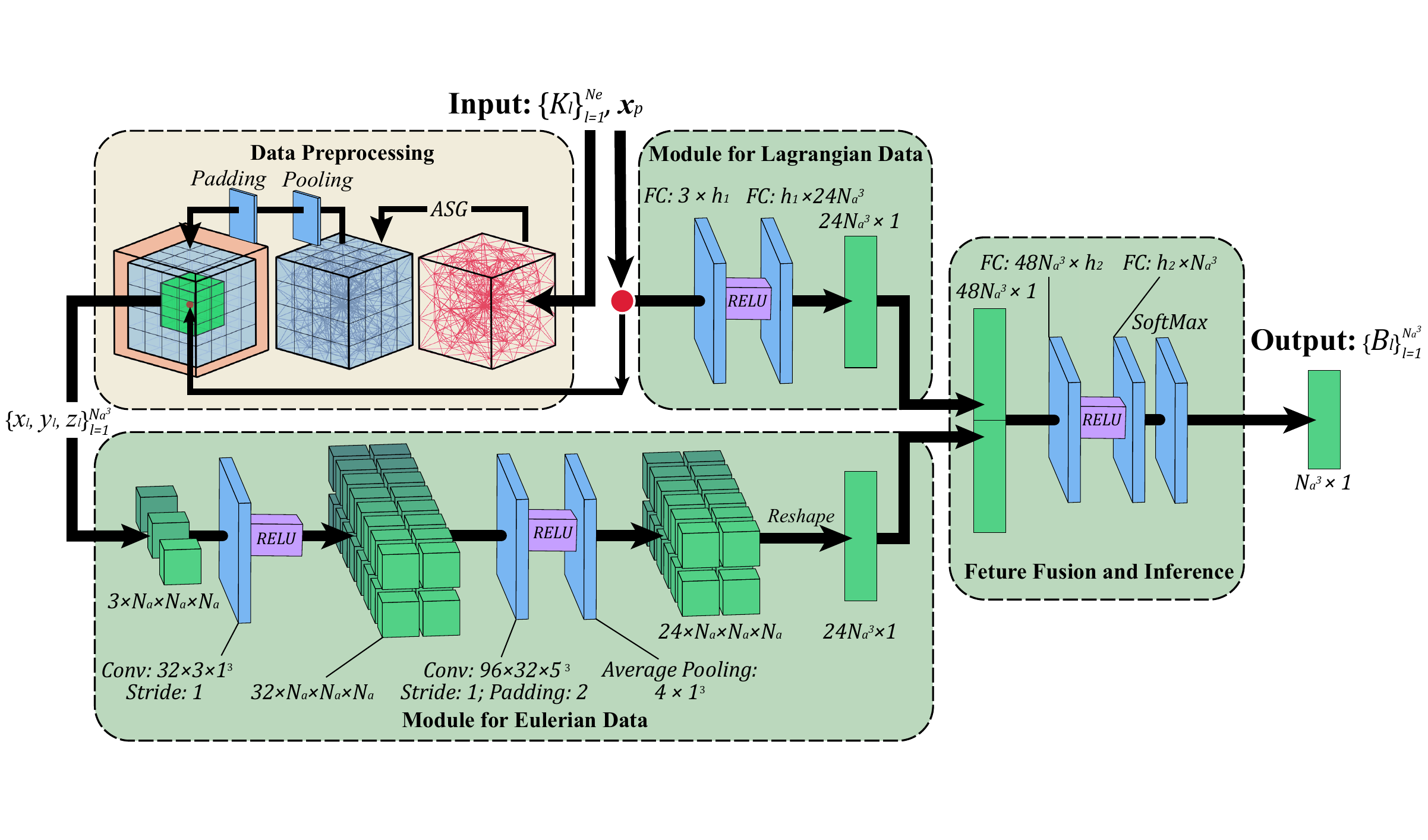}
  \caption{The prediction process of the CNN is shown, including the data preprocessing module (yellow block) and the CNN modules (green blocks). The configurations of fully connected (FC) layers and convolutional (Conv) layers are given in the figure, where $h_1$ and $h_2$ represent the dimensions of FC layers.}\label{fig:Fig2}
\end{figure}

Essentially, the CNN predicts $K_p$ by approximating the mapping from $\boldsymbol{x}_p$ to $\left\{K_l\right\}_{l=1}^{N_e}$. Therefore, it necessitates both $\boldsymbol{x}_p$ and $\left\{K_l\right\}_{l=1}^{N_e}$ as input to represent the spatial information from the Lagrangian and Eulerian reference frames, respectively. However, providing the entire $\left\{K_l\right\}_{l=1}^{N_e}$ as input would incur high complexity of both the prediction and computation, especially for high-resolution models. To lower the complexity, the following simplifications are applied: Firstly, elements data $\left\{K_l\right\}_{l=1}^{N_e}$ are represented using their barycentric coordinates, $\left\{x_l^*,y_l^*,z_l^*\right\}_{l=1}^{N_e}$, instead of the original mesh data, which are defined as follows:
\begin{equation}
  \left[\begin{array}{l}
    x_l^* \\
    y_l^* \\
    z_l^*
    \end{array}\right]=\frac{1}{4}\left(\left[\begin{array}{l}
  x_l^1 \\
  y_l^1 \\
  z_l^1
  \end{array}\right]+\cdots+\left[\begin{array}{l}
  x_l^4 \\
  y_l^4 \\
  z_l^4
  \end{array}\right]\right) . \quad \forall l \in 1, \cdots, N_e .\nonumber
\end{equation}
Secondly, the elements around $\boldsymbol{x}_p$ are filtered using the ASG method, which overlays a cubic Cartesian grid on the original mesh. To address the non-uniformity issue in the ASG method, an average pooling layer is subsequently applied, which normalizes the number of $K_l$ within each Cartesian element and allows the Eulerian data to be stored in a spatially organized data structure. Thus, denoting the filtered elements around $\boldsymbol{x}_p$ as $\left\{K_l\right\}_{l=1}^{N_a}$, where $N_a$ is the number of selected elements that satisfied $N_a\ll N_e$, these elements are filtered within a cubic region centered at $\boldsymbol{x}_p$.
\begin{equation}
  \left\{K_l\right\}_{l=1}^{N_a} \subset\left\{K_l\right\}_{l=1}^{N_e}.\nonumber
\end{equation}
$N_a$ is dynamically adjusted to ensure that $\boldsymbol{x}_p$ remains within the cubic region. To preserve the integrity of the filtered data at the boundaries, a padding layer is added after the pooling layer, extending $\left\{K_l\right\}_{l=1}^{N_e}$ at the boundary. Inputting $\left\{K_l\right\}_{l=1}^{N_a}$ instead of $\left\{K_l\right\}_{l=1}^{N_e}$ substantially reduces feature complexity, as illustrated in Figure~\ref{fig:Fig2}.

For the CNN's output, instead of directly outputting $K_t$, the original mapping problem is transformed into a multiclass classification problem to reduce prediction complexity. Specifically, the CNN outputs the probability distribution of $\boldsymbol{x}_p$ belonging to each $K_l \in \left\{K_l\right\}_{l=1}^{N_a}$, where the element with the highest probability is selected as $K_p$. The prediction of the CNN can be described as
\begin{equation}
  \label{equ:7}
  \left\{B_l\right\}_{l=1}^{N_a}=\operatorname{CNN}\left(\left\{x_l^*, y_l^*, z_l^*\right\}_{l=1}^{N_a},\boldsymbol{x}_p\right),
\end{equation}
\begin{equation}
  \label{equ:8}
  B_p=\max \left(\left\{B_l\right\}_{l=1}^{N_a}\right),
\end{equation}
where $B_l$ is the probability of $\boldsymbol{x}_p\in K_l$.

Within the CNN, to resolve the multidimensional input specified in \eqref{equ:7}, $\left\{x_l^*,y_l^*,z_l^*\right\}_{l=1}^{N_a}$ and $\boldsymbol{x}_p$ are initially processed in two separate modules. The first module consists of two convolutional layers and a pooling layer to process $\left\{x_l^*,y_l^*,z_l^*\right\}_{l=1}^{N_a}$. The first convolutional layer extracts spatial features from each $K_l\in\left\{K_l\right\}_{l=1}^{N_a}$, while the second layer captures the correlations among $\left\{K_l\right\}_{l=1}^{N_a}$. The second module processes $\boldsymbol{x}_p$ through two fully connected layers to capture its features. For feature fusion, the outputs of both modules are concatenated, followed by two fully connected layers to produce logits. These logits are then converted into a probability distribution over $\left\{K_l\right\}_{l=1}^{N_a}$ using a softmax layer \citep{martins2016softmax}:
\begin{equation}
  B_l=\frac{e^{H_l}}{\sum_{g=1}^{N_a} e^{H_g}}\nonumber
\end{equation}
where $\left\{H_l\right\}_{l=1}^{N_a}$ is the output of the last fully connected layer. Additionally, the ReLU activation function \citep{agarap2018deep} is employed between layers.

During the training phase, the SNS method is applied for particle tracking and collecting training datasets. Based on $K_t$ identified by the SNS method, the true labels $B_l^{'}$ in the training dataset is defined as
\begin{equation}
  B_l^{\prime}= \begin{cases}1 & \text { if } l=t \\ 0 & \text { if } l \neq t .\end{cases}\nonumber
\end{equation}
Based on $B_l^{'}$, the CNN quantifies the prediction error by the Cross-Entropy Loss Function, given as
\begin{equation}
  \operatorname{Loss}_p=-\sum_{l=1}^{N_a} B_l^{\prime} \log \left(B_l\right).\nonumber
\end{equation}
In addition, the Adam optimizer is used to optimize the convergence of training the CNN.

With $p$ determined by \eqref{equ:8}, \eqref{equ:6} can be transformed into
\begin{equation}
  K_t=\operatorname{SNS}\left(\boldsymbol{x}_p, K_p,\left\{K_l\right\}_{l=1}^{N_e},\left\{\left[m_1^l, \cdots m_4^l\right]\right\}_{l=1}^{N_e}\right).\nonumber
\end{equation}
The process of the CNN-SNS method is shown in Algorithm~\ref{alg:CNN-SNS}.
\begin{algorithm}[H]
  \caption{The CNN-SNS method}\label{alg:CNN-SNS}
  \begin{algorithmic}[1]
  \State \textbf{Input:} $\boldsymbol{x}_p, K_m, \{K_l\}_{l=1}^{N_e}, \{[m_1^l, \dots, m_4^l]\}_{l=1}^{N_e}$
  \State \textbf{Output:} $K_t$

  \State $\{[x_l^*, y_l^*, z_l^*]\}_{l=1}^{N_a} \gets {\rm ASG}(\{K_l\}_{l=1}^{N_e})$
  \State $K_l = K_p = {\rm CNN}(\{[x_l^*, y_l^*, z_l^*]\}_{l=1}^{N_a}, [\boldsymbol{x}_p, y_p, z_p])$
  \State $I_t \gets 0$
  \While{true}
      \State $I_t \gets I_t + 1$
      \State Calculate $\{\lambda_i^{K_l}\}_{i=1}^4$
      \If{$\{\lambda_i^{K_l}\}_{i=1}^4$ satisfy $C_l$}
          \State \textbf{break}
      \EndIf
      \State Find $j \in \{1, \dots, 4\}$ that satisfy $\lambda_j^{K_l} = \min(\{\lambda_i^{K_l}\}_{i=1}^4)$
      \If{$m_j^l \neq -1$}
          \State $K_l \gets m_j^l$
      \Else
          \State \textbf{break}
      \EndIf
  \EndWhile
  \State \Return $K_t \gets K_l$
  \end{algorithmic}
\end{algorithm}

\section{Numerical results}
\label{sec:3}

In this section, the CNN-SNS method is tested with the three-dimensional simulations on unstructured meshes, including the lid-driven cavity flow and the flow past a sphere.

\subsection{Lid-driven cavity flow problem}
\label{subsec:4}

\begin{figure}[t]
	\begin{minipage}[t]{0.47\linewidth}
    \centering
		\includegraphics[width=6cm,height=6cm]{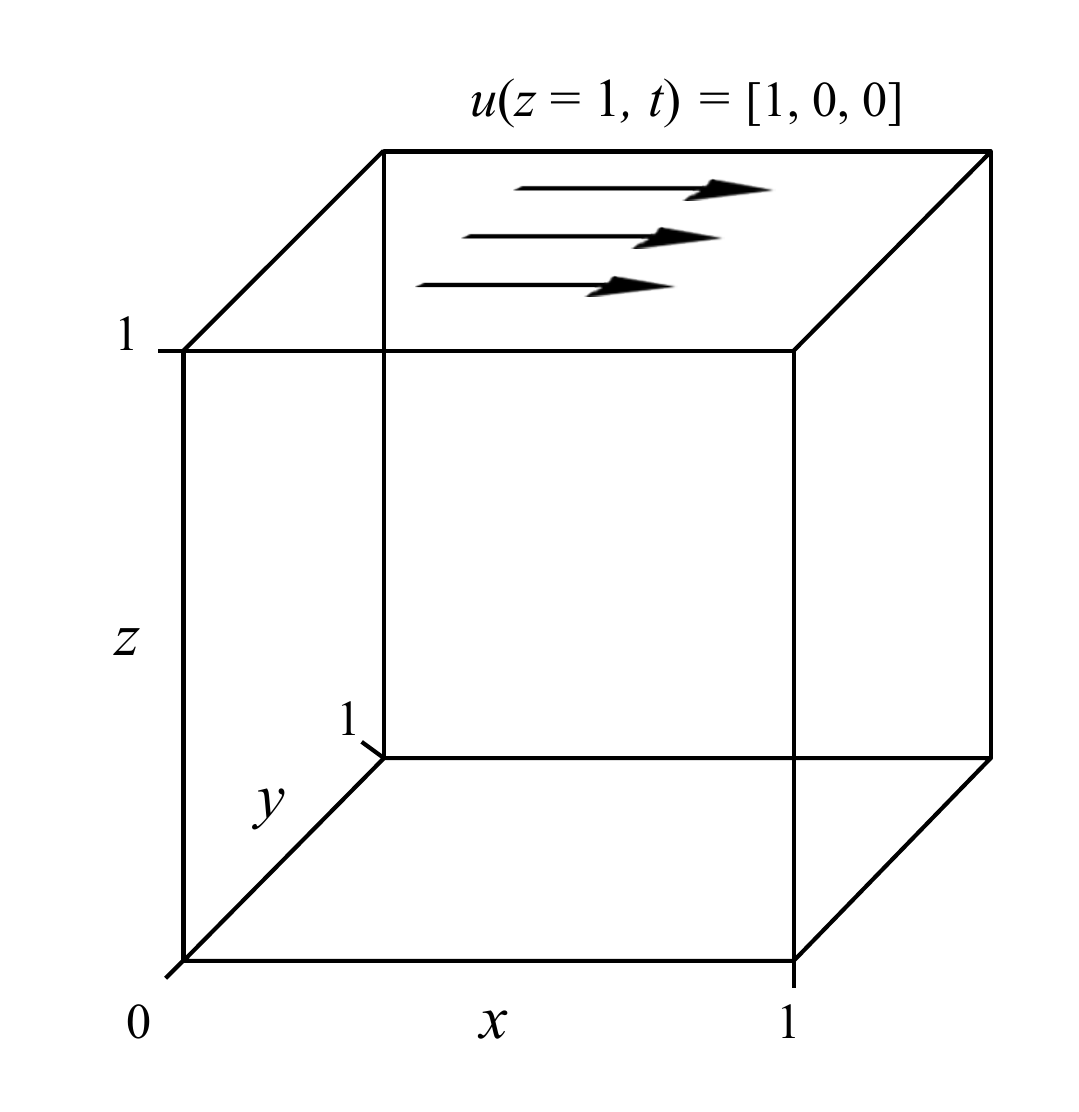}
		\caption{The model of the three-dimensional lid-driven cavity flow, with a velocity boundary on the upper surface where $\boldsymbol{u}(z=1,t)=[1,0,0]$, and solid wall boundaries on other surfaces.}\label{fig:Fig3}
	\end{minipage}
  \hspace{0.04\linewidth}
	\begin{minipage}[t]{0.47\linewidth}
    \centering
    \includegraphics[width=6cm,height=6cm]{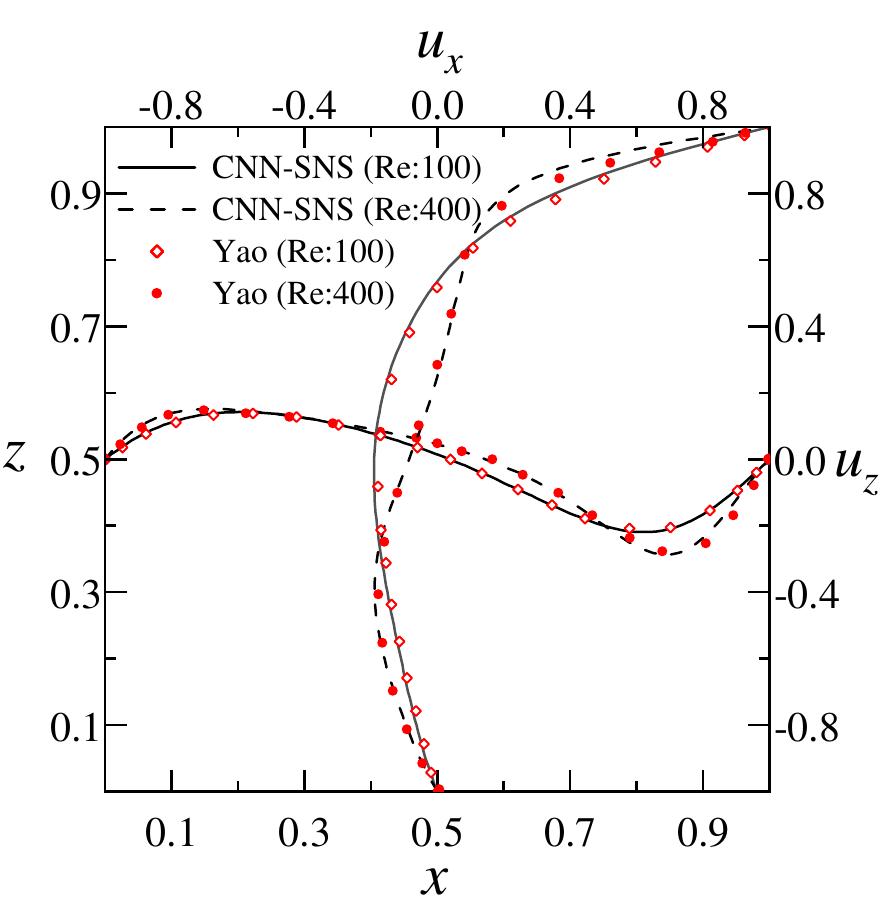}
    \caption{The velocity components along the centerlines $[x,0.5,0.5]$ and $[0.5,0.5,z]$ for $N_e= 1.123 \times 10^7$, $\mathrm{Re}=100,400$, compared to Yao \textit{et al.}'s result \citep{yao2010balancing}.}\label{fig:Fig4}
  \end{minipage}
\end {figure}

The configurations of the three-dimensional lid-driven cavity flow is shown in Figure~\ref{fig:Fig3}. The simulations are performed using the L-G solver provided by the ADVENTURE system \citep{ogino2005seismic}, with the Reynolds numbers of $\mathrm{Re} \in \left\lbrace 100, 400\right\rbrace$ and $\Delta t = 0.1\,\text{s}$ for $9\,\text{s}$. Three different mesh resolutions are included, where $N_e \in \left\lbrace 1.123 \times 10^7, 5.014 \times 10^7, 8.192 \times 10^7 \right\rbrace$. In the CNN-SNS method, we set $N_a = 59261$, $n_t = 3$, $h_1 = 256$, $h_2 = 512$. Both methods are tested using four Intel(R) Xeon(R) Platinum 8269CY CPUs, each with 26 cores.

Figure~\ref{fig:Fig4} compares the results of the CNN-SNS method $N_e = 1.123 \times 10^7$ with those of Yao \textit{et al.} \citep{yao2010balancing}. A close agreement has been achieved, demonstrating the reliability of the CNN-SNS method. For simplicity, the subsequent simulations are conducted at $\mathrm{Re} = 400$.

\begin{figure}[H]
  \centering
  \includegraphics[width=0.6\linewidth]{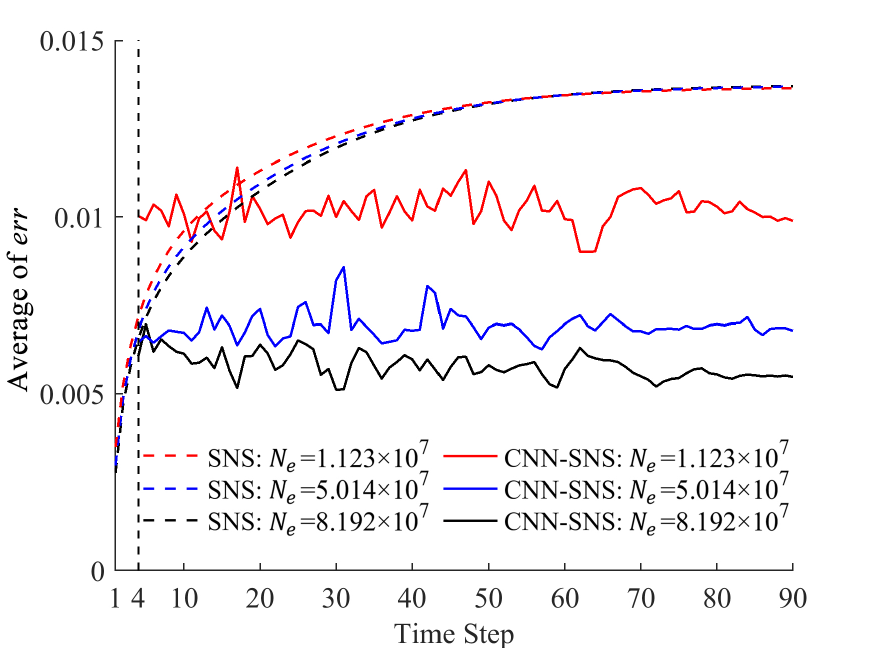}
  \caption{Average of $err$ in both the SNS and the CNN-SNS methods at each time step in the lid-driven cavity flow simulations.}\label{fig:Fig5}
\end{figure}

Figure~\ref{fig:Fig5} illustrates the error of the tracking origins for both methods at each time step. The error, denoted as $err$, is defined as the distance between the tracking origins and $K_t$.
\begin{equation}
  err=\left\{
  \begin{array}{ll}
  \left(\left(x_t^*-x_m^*\right)^2+\left(y_t^*-y_m^*\right)^2+\left(z_t^*-z_m^*\right)^2\right)^{0.5}, & \text{in SNS} \\
  \left(\left(x_t^*-x_p^*\right)^2+\left(y_t^*-y_p^*\right)^2+\left(z_t^*-z_p^*\right)^2\right)^{0.5}, & \text{in CNN - SNS}
  \end{array}
  \right.\nonumber
\end{equation}
As time goes on, the cavity flow is accelerated by the lid, resulting in an increase in both $|\boldsymbol{u}|$ and $err$ in the SNS method. For the CNN-SNS method, after training on data from the first $n_t$ time steps, its $err$ remains at a low level in the subsequent time steps, demonstrating the generalization capability of the CNN. Towards the end of the simulation, the flow field tends to stabilize, along with $err$ in both methods. For simulations with different $N_e$, $err$ remains largely constant in the SNS method, whereas it decreases as $N_e$ increases in the CNN-SNS method, indicating that the CNN yields higher accuracy as the model's resolution grows.

\begin{figure}[H]%% placement specifier
  \centering%% For centre alignment of image.
  \includegraphics[width=0.8\linewidth]{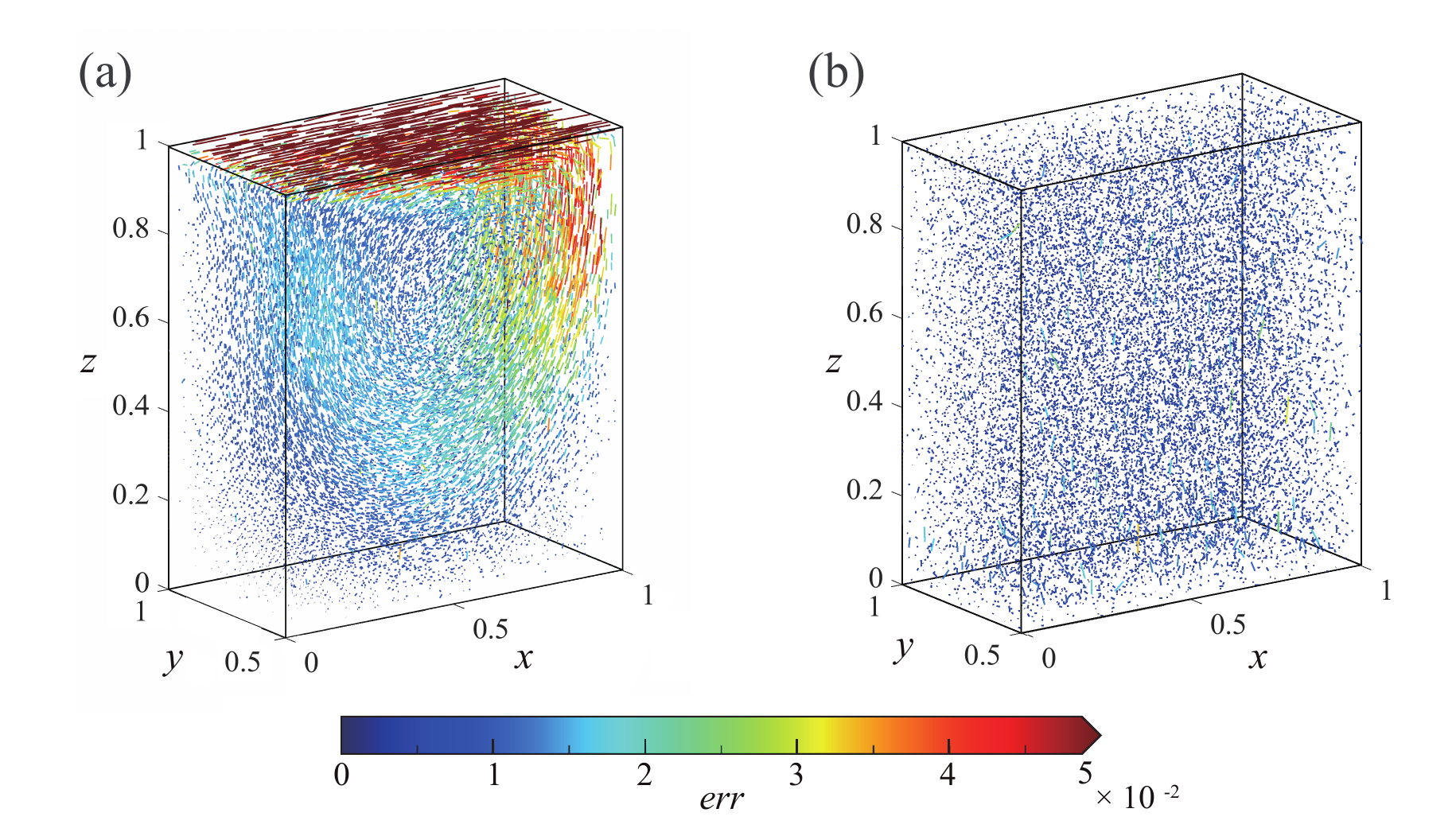}
  %% Use \caption command for figure caption and label.
  \caption{The distributions of $err$ and the tracking paths in the lid-driven cavity model of $N_e= 8.192 \times 10^7$ at $t = 9\,\text{s}$: (a) in the SNS method; (b) in the CNN-SNS method.}\label{fig:Fig6}
\end{figure}

Optimization in $err$ shortens the tracking path, as illustrated in Figure~\ref{fig:Fig6}. In the SNS method, the distribution of tracking paths aligns with the velocity distribution of the flow field, exhibiting a ring-like pattern. The tracking paths tend to be longer in regions with high $err$. In contrast, $err$ remains consistently low in the flow field in the CNN-SNS method, and the tracking paths exhibit a short, scattered distribution.

\begin{figure}[H]
	\begin{minipage}[t]{0.47\linewidth}
    \centering
		\includegraphics[width=6cm,height=5cm]{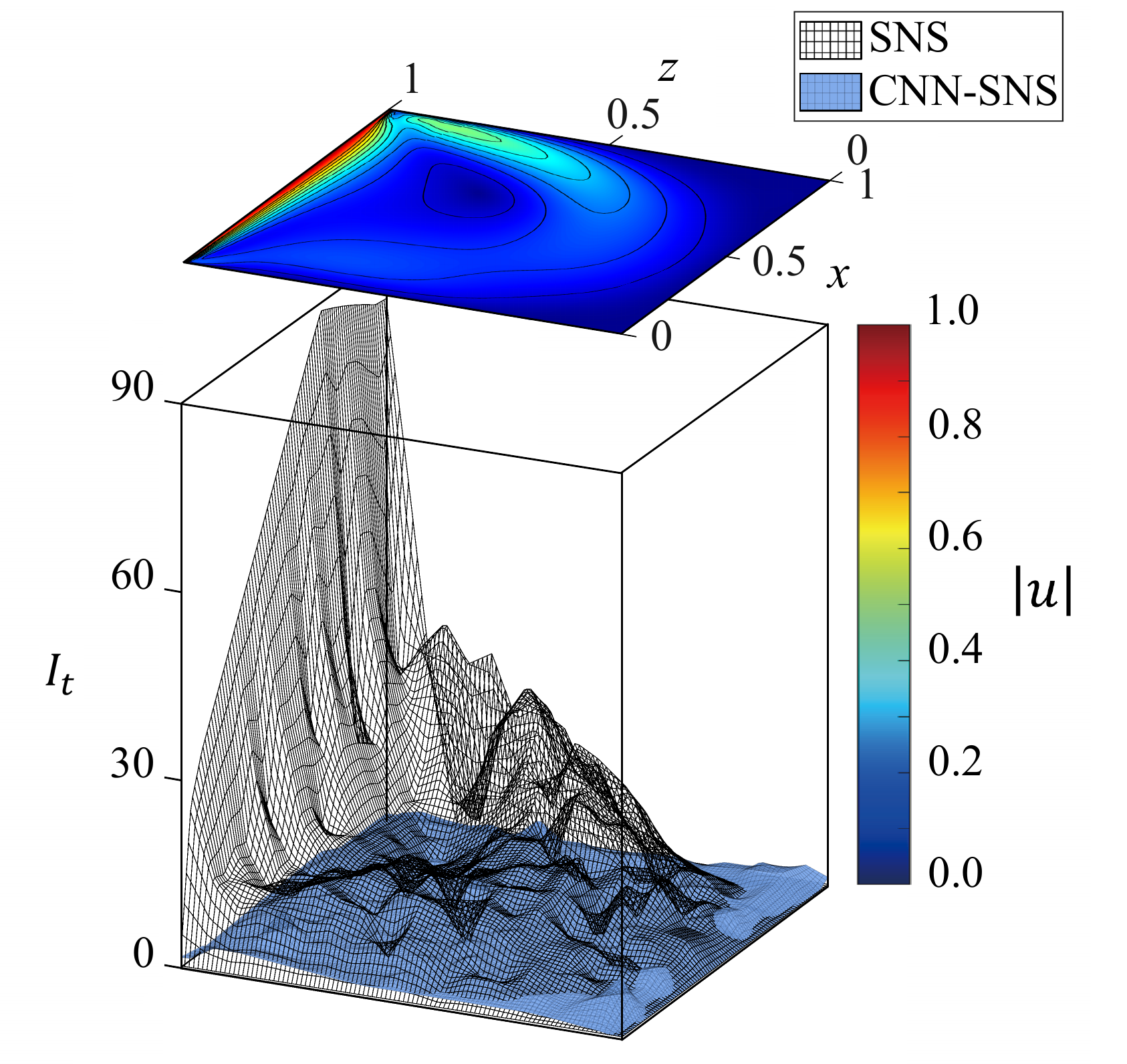}
		\caption{$|\boldsymbol{u}|$ and $I_t$ in both the SNS and the CNN-SNS methods on the mid-plane of the lid-driven cavity model where $y = 0.5$, for $N_e = 8.192 \times 10^7$ at $t = 9\,\text{s}$.}
    \label{fig:Fig7}
	\end{minipage}
  \hspace{0.04\linewidth}
	\begin{minipage}[t]{0.47\linewidth}
    \centering
    \includegraphics[width=6cm,height=5cm]{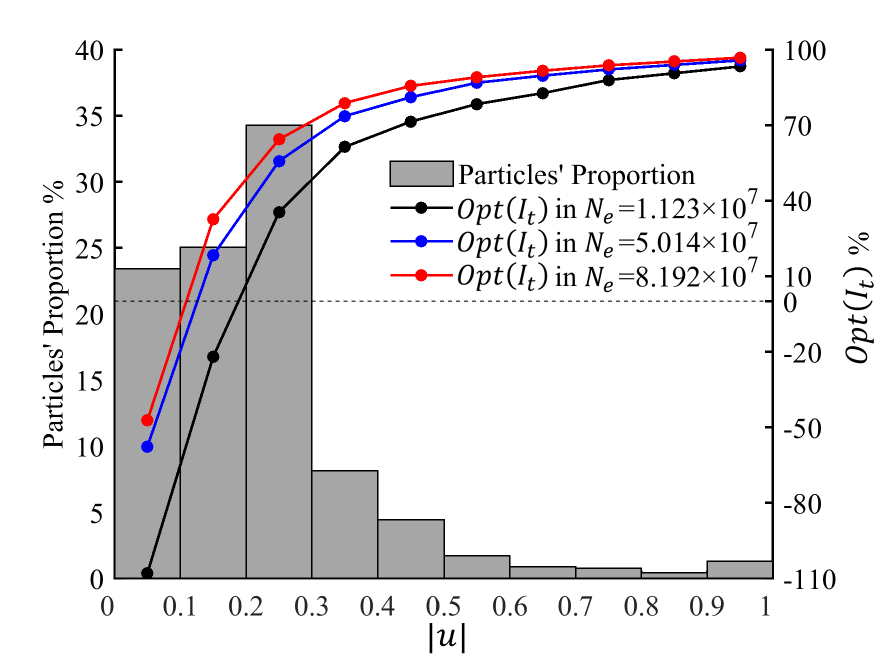}
    \caption{$Opt\left(I_t\right)$ and the proportion of particles with different $|\boldsymbol{u}|$ in the lid-driven cavity simulations for $N_e= 8.192 \times 10^7$ at $t=9\,\text{s}$.}
    \label{fig:Fig8}
  \end{minipage}
\end {figure}

Along the shorter tracking paths, the CNN-SNS method achieves targets with lower $I_t$. Figure~\ref{fig:Fig7} shows $I_t$ incurred by both methods in the mid-plane at $y = 0.5$. In the SNS method, $I_t$ is generally distributed corresponds to $|\boldsymbol{u}|$, except near the boundary at $z=1$. This occurs because the tracking toward the high-velocity particles near $z=1$ terminates at $x=0$ in advance. In contrast, the CNN-SNS method exhibits a more stable $I_t$, consistently remaining at a low level. Notably, $I_t$ of the CNN-SNS method exceeds that of the SNS method only in the low-velocity region around $z=0$, resulting in a minor reduction in optimization efficiency. Figure~\ref{fig:Fig8} presents the positive correlation between $Opt\left(I_t\right)$ and $|\boldsymbol{u}|$, indicating that the CNN-SNS method rapidly reaches a positive optimization as $|\boldsymbol{u}|$ increases from zero.

The statistics of the computational consumption in both methods throughout the simulations are summarized in Table~\ref{tab:Tab1}. The optimization rates of the CNN-SNS method over the SNS method are given by
\begin{equation}
  Opt\left(C\right)=\frac{C\left(\text{SNS}\right) - C\left(\text{CNN-SNS}\right)}{C\left(SNS\right)}\nonumber
\end{equation}
where $Opt\left(C\right)$ denotes the optimization rate for the metric $C$, which could be computational time or $I_t$. In the SNS method, both computational time and $I_t$ increase with $N_e$, as both the tracking complexity and the number of the particles are positively correlated with $N_e$. For the CNN-SNS method, the number of particles processed by the CNN increased with $N_e$, resulting in an increasing CNN inference time, which remains manageable compared to the SNS method. Thus, the total time cost in the CNN-SNS method is significantly lower than that of the SNS method, and the gap widens as $N_e$ increases, resulting in an increasing optimization of time costs.

\begin{table}[H]
  \centering
  \caption{The computational consumption of the SNS and the CNN-SNS methods in the lid-driven cavity flow simulations.}\label{tab:Tab1}
  \begin{tabular}{@{}cccc@{}}
  \toprule
  {$N_e$} & $1.123 \times 10^7$ & $5.014 \times 10^7$ & $8.192 \times 10^7$\\
  \addlinespace
  \multicolumn{4}{c}{SNS} \\
  \midrule
  Average of $err$ & 0.0125 & 0.0123 & 0.0123 \\
  Average of $I_t$ & 5.002 & 8.801 & 11.561 \\
  Time cost (s) & 7.087 & 65.560 & 137.233 \\
  \addlinespace
  \multicolumn{4}{c}{CNN-SNS} \\
  \midrule
  Average of $err$ & 0.0102 & 0.0069 & 0.0058 \\
  Average of $I_t$ & 4.193 & 4.180 & 4.305 \\
  Time cost in CNN (s) & 5.075 & 30.367 & 64.103 \\
  Time cost in SNS (s) & 1.952 & 5.017 & 8.711 \\
  Time cost (s) & 7.027 & 35.384 & 72.814 \\
  \addlinespace
  \multicolumn{4}{c}{$Opt$} \\
  \midrule
  Average of $err$ & 18.400\% & 43.902\% & 52.846\% \\
  Average of $I_t$ & 16.173\% & 52.505\% & 62.763\% \\
  Time cost (s)& 0.846\% & 46.028\% & 46.941\% \\

  \bottomrule
  \end{tabular}
\end{table}

\begin{figure}[H]%% placement specifier
  \centering%% For centre alignment of image.
  \includegraphics[width=1\linewidth]{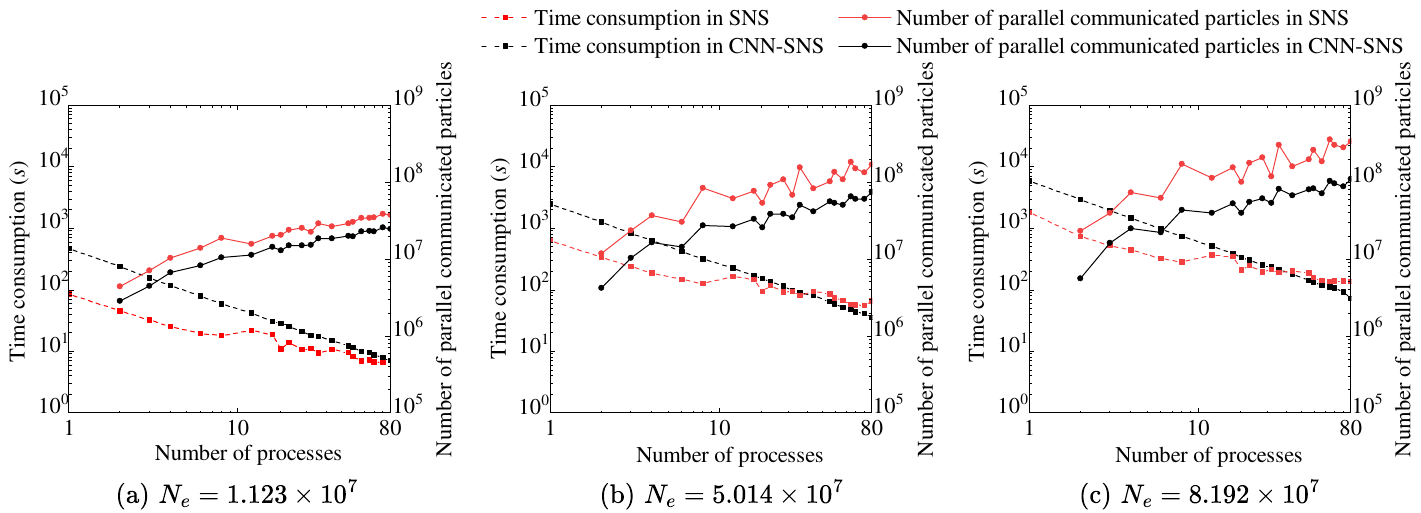}
  %% Use \caption command for figure caption and label.
  \caption{The time consumptions of the SNS and the CNN-SNS methods applied in different parallel configurations, and the number of parallel communicated particles in both methods.}
  \label{fig:Fig9}
\end{figure}

In other parallel configurations, the computational costs of both methods are shown in Figure~\ref{fig:Fig9}. In a single process, the CNN-SNS method incurs higher time costs than the SNS method due to the inference procedure of the CNN. As the level of parallelization increases, the number of particles being communicated inter-process in the CNN-SNS method remains consistently lower than that in the SNS method, indicating a superior parallel scalability for the CNN-SNS method. Thus, as the number of processes approaches $80$, the performance of the SNS method degrades, and the CNN-SNS method achieves a positive optimization in time costs. 

Simulations in the lid-driven cavity flow problem demonstrate that the CNN-SNS method can achieve considerable improvements in computational efficiency even in models dominated by low-speed particles, as illustrated in Figure~\ref{fig:Fig8}. To further evaluate the performance of the CNN-SNS method in a high-velocity flow field, the simulation of the flow around a sphere model is conducted.

\subsection{Flow around a sphere problem}
\label{subsec:5}

\begin{figure}[H]
	\begin{minipage}[t]{0.47\linewidth}
    \centering
		\includegraphics[width = 1\linewidth]{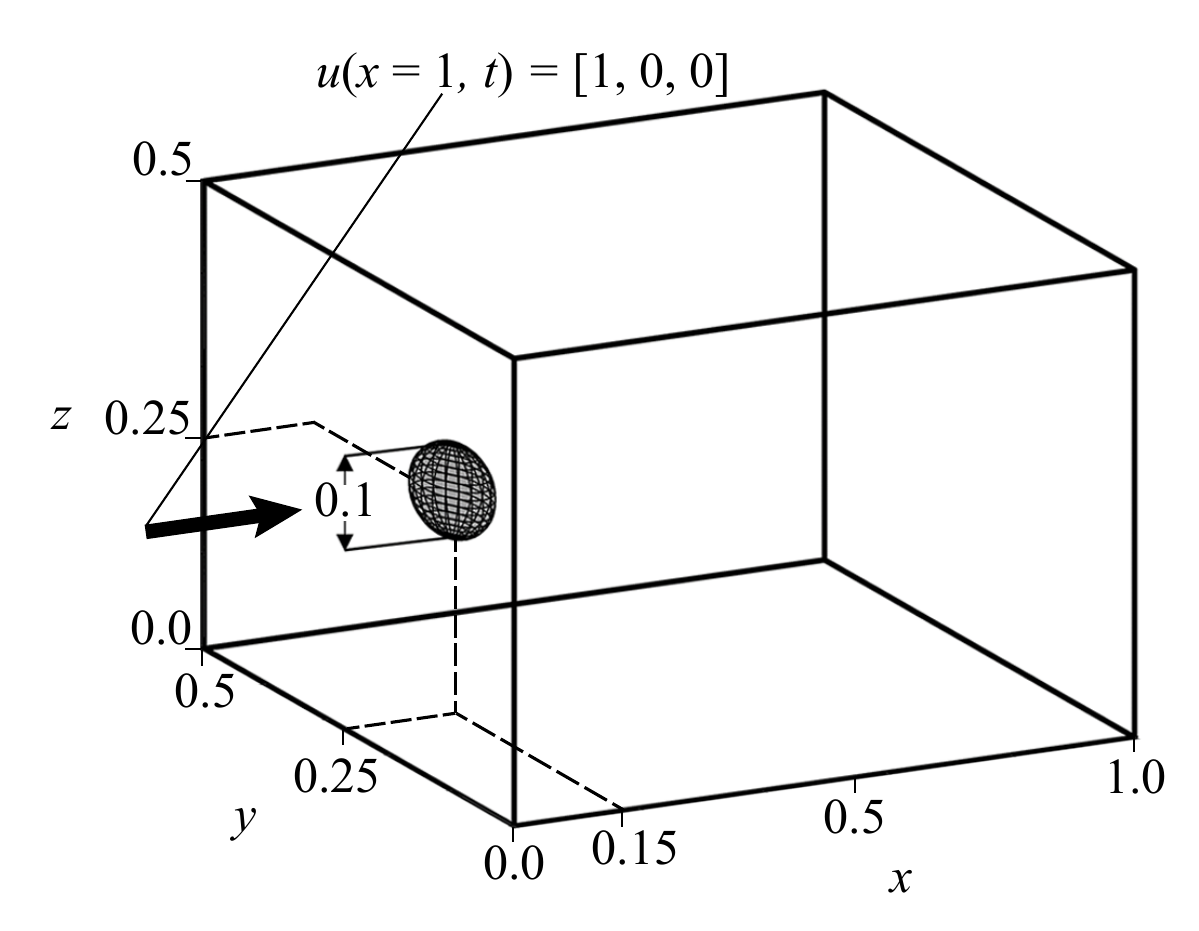}
		\caption{The model of the three-dimensional flow around a sphere, with the sphere of radius $0.05$ in $[0.15,0.25,0.25]$. It contains a solid wall boundary on the sphere and the other surfaces.}
    \label{fig:Fig10}
	\end{minipage}
  \hspace{0.04\linewidth}
	\begin{minipage}[t]{0.47\linewidth}
    \centering
    \includegraphics[width = 1\linewidth]{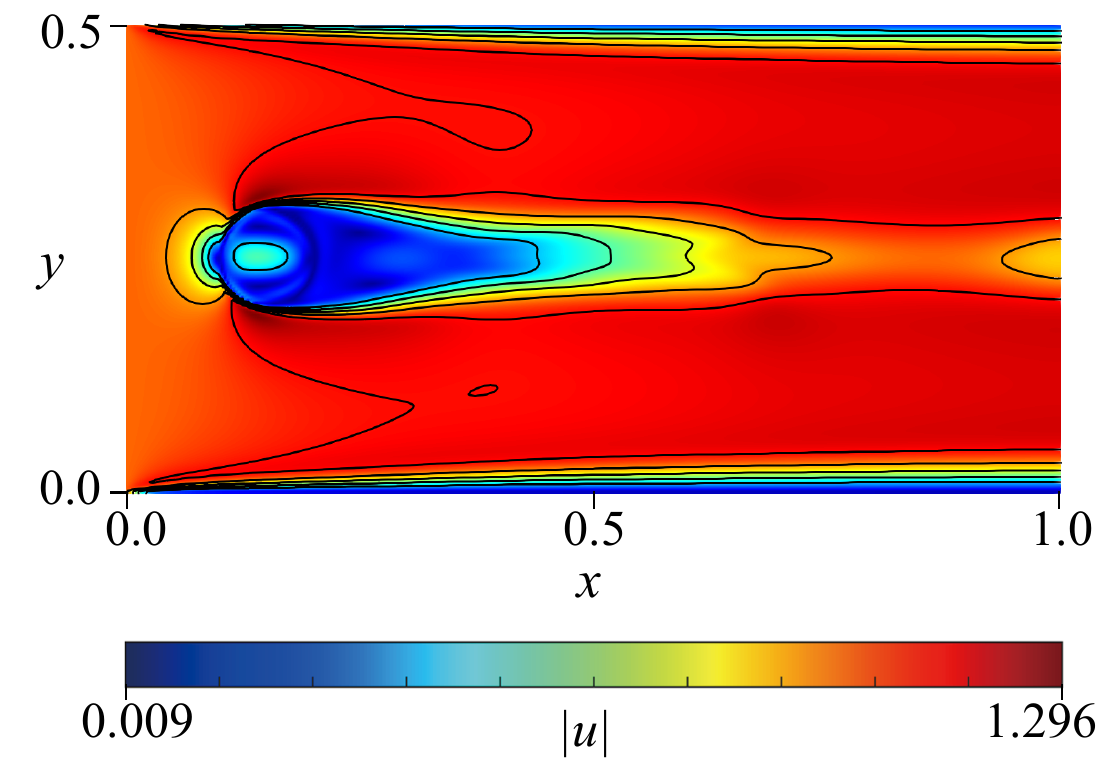}
    \caption{The distribution of $|\boldsymbol{u}|$ on the mid-plane of $z = 0.25$, in the flow around a sphere of $N_e=9.176 \times 10^6$ at $t = 20\,\text{s}$.}
    \label{fig:Fig11}
  \end{minipage}
\end {figure}

The configurations of the flow around a sphere is depicted in Figure~\ref{fig:Fig10}, with $\Delta t = 0.1\, \text{s}$, $N_e \in \left\lbrace 9.176 \times 10^6, 5.014 \times 10^7, 9.789 \times 10^7\right\rbrace$, and $\mathrm{Re} = 400$ for $20\,\text{s}$. In the CNN-SNS method, we set $N_a = 59261$, $n_t=3$, $h_1 = 256$, $h_2 = 512$.

\begin{figure}[H]
	\begin{minipage}[b]{0.47\linewidth}
    \centering
		\includegraphics[width=1\linewidth]{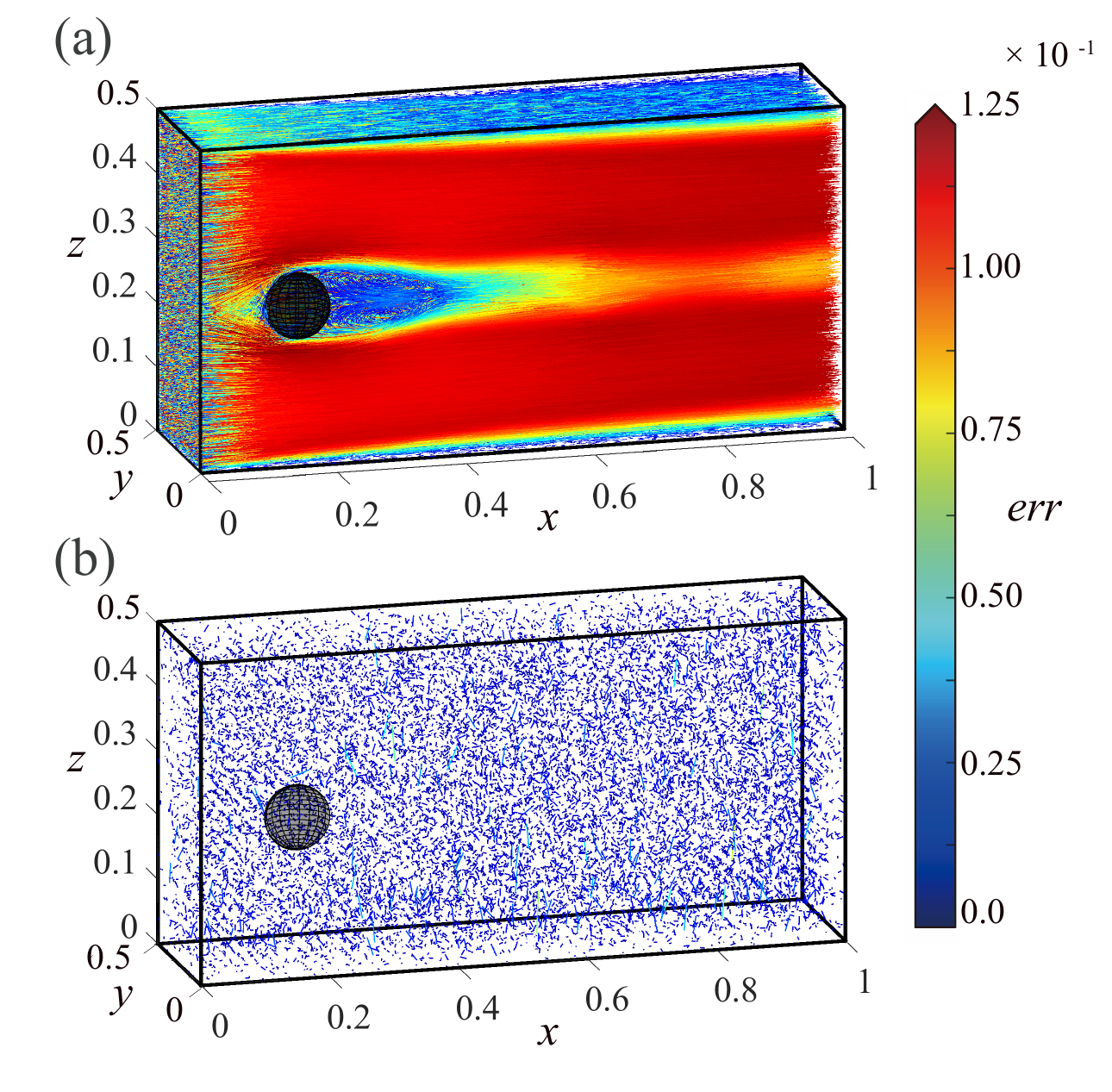}
		\caption{The distributions of $err$ and the tracking paths in the model of $N_e= 9.789 \times 10^7$ at $t = 20\,\text{s}$: (a) in the SNS method; (b) in the CNN-SNS method.}
    \label{fig:Fig12}
	\end{minipage}
  \hspace{0.04\linewidth}
	\begin{minipage}[b]{0.47\linewidth}
    \centering
    \includegraphics[width=1\linewidth]{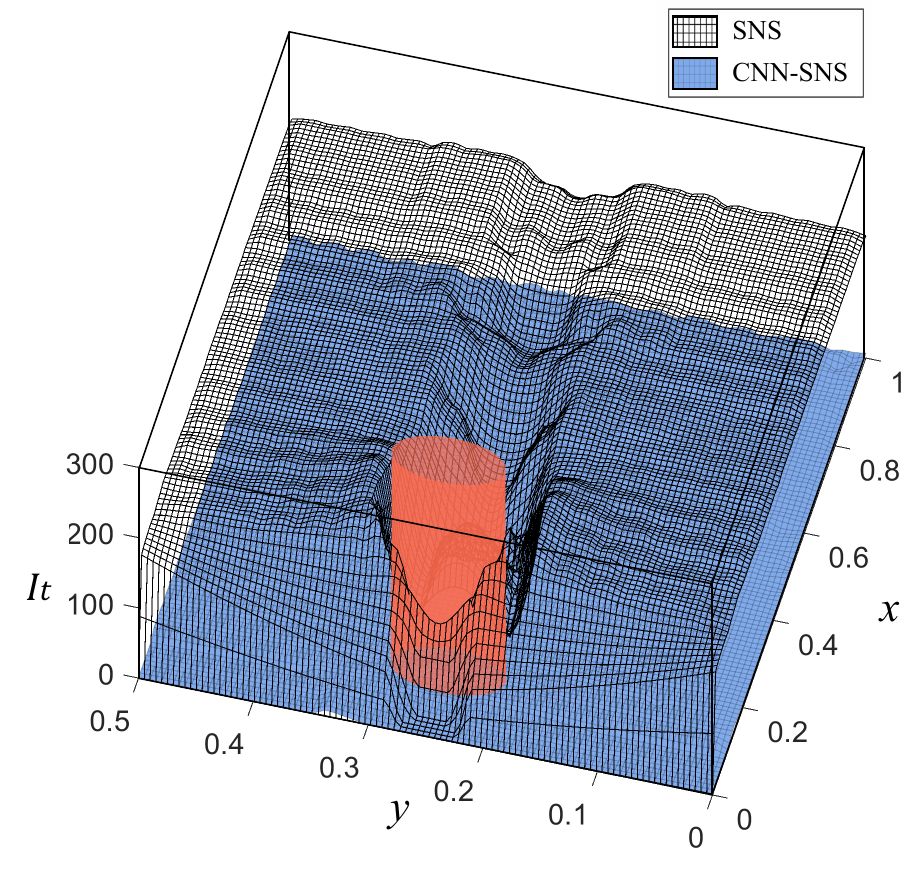}
    \caption{The distributions of $I_t$ in both the SNS and the CNN-SNS methods on the mid-plane of $z = 0.5$ in the flow around a sphere model for $N_e= 9.789 \times 10^7$ at $t = 20\,\text{s}$.}
    \label{fig:Fig13}
  \end{minipage}
\end {figure}

Figure~\ref{fig:Fig11} shows that the flow around a sphere exhibits high velocity, except behind the sphere and around the boundary. Consequently, the SNS method incurs high $err$ throughout the flow field and requires lengthy tracking paths to reach the targets, as shown in Figure~\ref{fig:Fig12}. In contrast, the CNN-SNS method still maintains a low $err$, enabling shorter tracking paths. The comparison of $I_t$ between the two methods is illustrated in Figure~\ref{fig:Fig13}. In the SNS method, $I_t$ follows the distribution of $|\boldsymbol{u}|$ and presents a wave-like pattern downstream of the sphere. In the CNN-SNS method, $I_t$ remains consistently low in the flow field.

The consumptions of both methods throughout the simulations are provided in Table~\ref{tab:Tab2}. The SNS method incurs significant computational time due to the high $err$ and the lengthy tracking paths. In the CNN-SNS method, $err$ and $I_t$ are comparable to those in the cavity flow problem. While the CNN remains at a similar scale compared to that in the cavity flow simulations, its time costs in the inference procedure increase due to the longer time scales in simulations. The time cost optimization remains consistently high across models with different resolutions. Comparing Table~\ref{tab:Tab2} and Table~\ref{tab:Tab1} reveals that the CNN-SNS method achieves greater performance gains in higher-velocity flow fields.

\begin{table}[H]
  \centering
  \caption{The computational consumption of the SNS and the CNN-SNS methods, in the flow around a sphere simulations with different $N_e$, and the optimization rate of the CNN-SNS method in $T$ and $\bar{I_t}$ compared to the SNS method.}\label{tab:Tab2}
  \begin{tabular}{@{}cccc@{}}
      \toprule
      {$N_e$} & $9.716 \times 10^6$ & $5.575 \times 10^7$ & $9.789 \times 10^7$\\
      \addlinespace
  \multicolumn{4}{c}{SNS} \\
  \midrule
  Average of $err$ & 0.1035 & 0.1032 & 0.1034 \\
  Average of $I_t$ & 62.372 & 103.006 & 124.829 \\
  Time cost (s) & 538.126 & 1902.96 & 8405.214 \\
  \addlinespace
  \multicolumn{4}{c}{CNN-SNS} \\
  \midrule
  Average of $err$ & 0.0072 & 0.0055 & 0.0048 \\
  Average of $I_t$ & 5.132 & 4.916 & 5.225 \\
  Time cost in CNN (s) & 12.059 & 68.143 & 228.945 \\
  Time cost in SNS (s) & 7.359 & 14.400 & 22.916 \\
  Time cost (s) & 19.418 & 82.543 & 251.861 \\

  \addlinespace
  \multicolumn{4}{c}{$Opt$} \\
  \midrule
  Average of $err$ & 93.044\% & 94.671\% & 95.358\% \\
  Average of $I_t$ & 91.772\% & 95.228\% & 95.814\% \\
  Time cost (s) & 96.392\% & 95.662\% & 97.004\% \\
  \bottomrule
  \end{tabular}
\end{table}

\begin{figure}[H]%% placement specifier
  \centering%% For centre alignment of image.
  \includegraphics[width=1\linewidth]{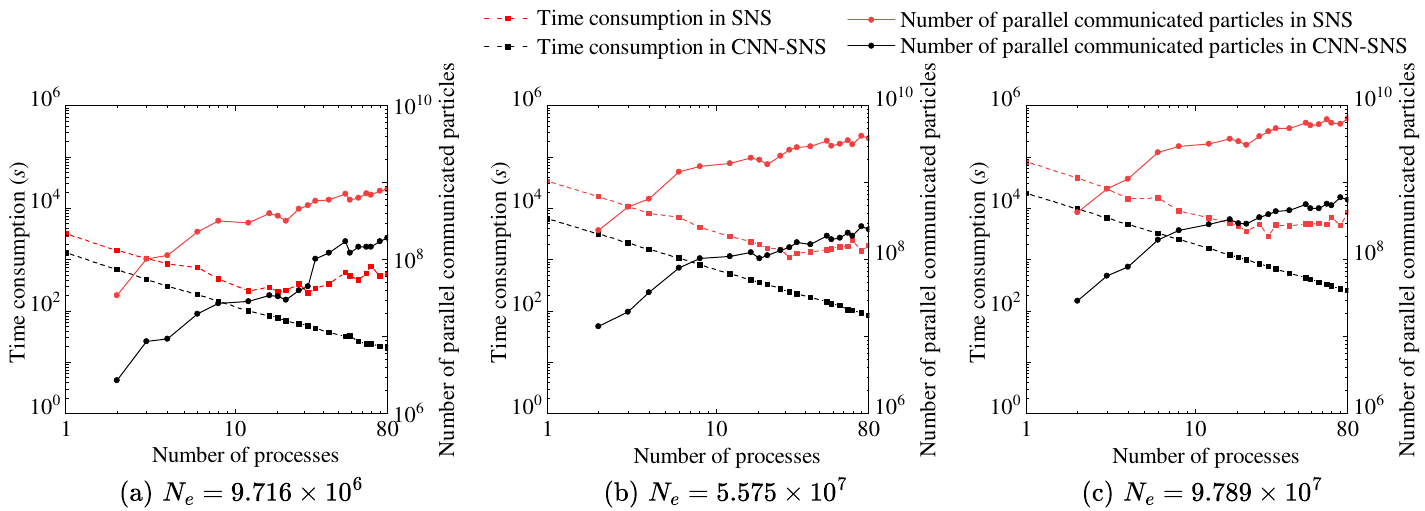}
  %% Use \caption command for figure caption and label.
  \caption{The time consumption of the SNS and the CNN-SNS methods applied in different parallel configurations, and the number of parallel communicated particles in both methods.}
  \label{fig:Fig14}
\end{figure}

Figure~\ref{fig:Fig14} provides the computational consumptions in different parallel configurations. The time costs of the CNN-SNS method are consistently lower than those of the SNS method, and the gap between the two methods widens as the number of processes increases. Additionally, comparing Figure~\ref{fig:Fig14} and Figure~\ref{fig:Fig9} indicates that the optimization in parallel scalability is superior in the high-velocity flow field.

\section{Conclusions}
\label{sec:4}
In this paper, the CNN-SNS method is developed to improve the efficiency of the particle tracking process in L-E approaches for large-scale flow simulations. Simulations demonstrate that the CNN-SNS method significantly reduces the computational consumption of the tracking process, with an acceptable cost in the CNN inference procedure. As the resolution of the Eulerian model grows, the CNN-SNS method achieves higher optimization in computational efficiency compared to the SNS method, especially in high-velocity flow fields. Furthermore, while the efficiency of the SNS method degrades as the parallelization scales up, the CNN-SNS method reduces the inter-processor communication demand substantially and exhibits superior parallel scalability. Overall, these findings demonstrate the advantage of the CNN-SNS method in large-scale parallel simulation.

So far, the tests have been conducted under cases with relatively limited parallel scales and low-speed flow conditions. In future work, we plan to further evaluate the performance of the CNN-SNS method in larger-scale, high-speed flow fields.

\section*{Acknowledgment}
\label{sec:5}

This work has been supported by the National Natural Science Foundation of China (NSFC) under Grant No. 11972384, the Guangdong Basic and Applied Basic Research Foundation - Guangdong-Hong Kong-Macao Applied Mathematics Center Project under Grant No. 2021B1515310001, and the Guangdong Basic and Applied Basic Research Foundation - Regional Joint Fund Key Project under Grant No. 2022B1515120009. Additionally, we extend our appreciation to the National Key Research and Development Program under Grant No. 2020YFA0712502 for its invaluable support in this research.

\section*{Declaration of generative AI and AI-assisted technologies in the writing process}
\label{sec:6}

During the preparation of this work the authors used ChatGPT in order to improve language and readability. After using this tool, the authors reviewed and edited the content as needed and take full responsibility for the content of the publication.

\bibliographystyle{elsarticle-num}
\bibliography{Library}

\end{document}